%% file: main.tex
\documentclass[lettersize,journal]{IEEEtran}
\usepackage{arydshln}
\usepackage{tikz}
\usepackage{physics,amsmath,amssymb,amsfonts}
\usepackage{algorithm}
\usepackage{algpseudocode}
\usepackage{soul}
\usepackage{array}
\usepackage{filecontents}
\usepackage{makecell}
\usepackage{booktabs} 
\usepackage{multirow} 
\usepackage[table]{xcolor}
\usepackage{graphicx}
\usepackage{subcaption}
\usepackage[normalem]{ulem}
\usepackage{url}
\newcommand{\ND}{{\textit{NetDiffuser}}}
\newcommand{\PF}{{\textit{Discrete}}}
\newcommand{\SF}{{\textit{Relative}}}

\begin{document}

\title{NetDiffuser: Deceiving DNN-Based Network Attack 
Detection Systems with Diffusion-Generated Adversarial Traffic}


\author{
Pratyay Kumar, Abu Saleh Md Tayeen, Satyajayant Misra, Huiping Cao, Jiefei Liu, Qixu Gong, and Jayashree Harikumar
\thanks{Pratyay Kumar, Satyajayant Misra, Huiping Cao, Jiefei Liu, and Qixu Gong are with the Department of Computer Science, New Mexico State University, Las Cruces, NM, USA (e-mail: \{pratyay, misra, hcao, jiefei, qixugong\}@nmsu.edu).}%
\thanks{Abu Saleh Md Tayeen is with the University of Hartford, CT, USA (e-mail: tayeen@hartford.edu).}%
\thanks{Jayashree Harikumar is with DEVCOM Analysis Center, WSMR, NM, USA (e-mail: jayashree.harikumar.civ@army.mil).}%
}



\maketitle

\input{abstract}
\input{introduction}
\input{related-work}
\input{modelsandassumptions}
\input{methodology}
\input{evaluation}
\input{conclusion}
\bibliographystyle{plain}
\bibliography{ref}
\end{document}

%% file: abstract.tex
\begin{abstract}
Deep learning (DL)-based Network Intrusion Detection System (NIDS) has demonstrated great promise in detecting malicious network traffic. However, they face significant security risks due to their vulnerability to adversarial examples (AEs). Most existing adversarial attacks 
maliciously perturb data to maximize misclassification errors. 
Among AEs, natural adversarial examples (NAEs) are particularly difficult to detect because they closely resemble real data, making them challenging for both humans and machine learning models to distinguish from legitimate inputs. 
%
%
Creating NAEs is crucial for testing and strengthening NIDS defenses. 
This paper proposes \ND\footnotemark[1], a novel framework for generating NAEs capable of deceiving NIDS.
%
\ND~consists of two novel components. First, a new  feature categorization algorithm is designed to identify relatively independent features in a network traffic. Perturbing these features minimizes changes while preserving network flow validity. The second component is a novel application of diffusion models to inject semantically consistent perturbations for generating NAEs. 
\ND~performance was extensively evaluated using three benchmark NIDS datasets across various model architectures and state-of-the-art adversarial detectors. 
Our experimental results show that \ND~achieves up to a 29.93\% higher attack success rate and reduces AE detection performance by at least 0.267 (in some cases up to 0.534) in the Area under the Receiver Operating Characteristic Curve (AUC-ROC) score compared to the baseline attacks.



\footnotetext[1]{\ND~is open source and available at \url{https://anonymous.4open.science/r/NetDiffuser-15B2/README.md}}
\end{abstract}



%% file: introduction.tex
\section{Introduction}
\label{sec:intro}
Deep learning (DL) technologies have achieved significant success across multiple tasks in various domains, including computer vision~\cite{du2022elements, zaidi2022survey} and natural language processing~\cite{lauriola2022introduction}.
Inspired by this success, and given its minimal reliance on feature engineering and the decreasing cost of parallel processing hardware, deep learning has also been adopted in the networking domain~\cite{zhang2019deep}, 
particularly for building
Network Intrusion Detection System (NIDS).
An NIDS is designed to identify malicious traffic to safeguard computers, networks, and data from attacks, unauthorized access, or alterations~\cite{ferrag2020deep}. As cyber attacks continue to evolve and traditional 
signature and rule-based intrusion detection methods such as
Snort~\cite{snort10.5555/1039834.1039864} lose their effectiveness, researchers have shown that Deep Neural Network (DNN)-based NIDSs can achieve significantly higher detection accuracy~\cite{naseer2018enhanced,sun2020dl,vinayakumar2019deep}.

Unfortunately, DNNs have also been shown to be vulnerable to adversarial examples (AE)~\cite{szegedy2013intriguing}, where subtle, imperceptible perturbations introduced into 
training samples can lead to wrong decisions made by the DL model. 
In recent years, adversarial attacks utilizing AEs have been effectively applied across multiple domains, including computer vision~\cite{khamaiseh2022adversarial}, natural language processing~\cite{dong2022adversarial}, and speech recognition~\cite{esmaeilpour2022towards}. 
%
In a white-box setting, where the architecture and gradients of a DL model 
are accessible, adversarial attacks can substantially increase a model's misclassification rates.
Given the crucial role of NIDS in defending critical infrastructure of individuals, enterprises, and governments against cyber attacks, adversarial attacks on the DL-based NIDS model pose a serious risk. Malicious traffic representing Denial-of-Service (DoS),
SQL injection, or man-in-the-middle attacks, can potentially be disguised as benign, evading detection by even the most accurate DL-based NIDS, thereby compromising their effectiveness and the security of their systems~\cite{clements2021rallying, hashemi2019towards}.


In this context, several researchers have applied traditional adversarial example (AE) generation methods~\cite{goodfellow2014explaining, kurakin2018adversarial, PGDmadry2018towards} that were initially developed for computer vision tasks, to demonstrate the vulnerability of DL-based NIDSs to such attacks. These studies often made simplistic assumptions about the problem and focused on the entire feature space~\cite{ibitoye2019analyzing, martin2019footprint, wang2018deep}. However, unlike in computer vision, where features are largely independent and can be modified freely, network-flow data is subject to strict domain-specific constraints~\cite{ConstDomain:journals/corr/abs-2011-01183}. 
These constraints ensure that any perturbations introduced into the data remain within the bounds of legitimate behavior. For instance, features such as `port number', `destination address', and `network service' must be valid and consistent in any generated AE to preserve its functional plausibility.
Any deviation from these feature constraints can lead to network flows that exhibit non-compliant or logically inconsistent behavior, such as violating protocol standards or containing contradictory feature values, which would likely be flagged as anomalous or adversarial by an NIDS.

While adherence to these constraints is critical for generating credible adversarial examples, current research often fails to account for them comprehensively. Many existing studies either entirely overlooked~\cite{ibitoye2019analyzing,wang2018deep} or partially implemented~\cite{teuffenbach2020subverting} domain-specific feature limitations. This oversight 
undermines the validity of the experimental setting and also risks overestimating the vulnerability of NIDS to adversarial threats.
Thus, some researchers~\cite{sheatsley2022adversarial, teuffenbach2020subverting} have extended AE 
crafting algorithms by incorporating domain-specific constraints tailored to NIDS. These approaches introduced perturbations to a carefully selected subset of features, ensuring that the semantics of traffic flows remain unchanged. However, the selection of these features relies heavily on domain expertise and involves manual interventions~\cite{sheatsley2022adversarial}, making it neither practical nor efficient for real-time attacks. 


Existing adversarial example (AE) detectors~\cite{Manda9709532} in the NIDS domain are primarily designed to identify carefully crafted adversarial examples that manipulate the model's decision boundary, enabling them to detect such flows relatively easily. 
Most existing AE detectors did not explore their defense strength again a new type of recently proposed adversarial samples, Natural Adversarial Examples (NAEs)~\cite{Chen_2023_ICCV}. NAEs exploit inherent variability within natural data distributions by introducing subtle deviations that resemble legitimate fluctuations. Despite that NAEs were first used to attack computer vision models, such samples, if created, may be utilized to attack DL models in other domains such as NDS.
To strengthen NIDS defenses, understanding the mechanics to create NAEs and being able to generate NAEs are important to test NIDS. However, the impact of NAEs on existing NIDS remains underexplored.

This paper presents a newly designed framework, \ND, to generate NAEs targeting DL-based NIDS.
The design of \ND~aims at tackling two key challenges. 
First, there is a lack of a systematic approach for identifying modifiable features that preserve both the functional integrity and statistical legitimacy of network traffic flows.
Altering such features requires strict adherence to domain constraints (e.g., protocol compliance, temporal consistency) to ensure that adversarial examples remain indistinguishable from benign network traffic.
Second, unlike traditional adversarial example studies, which primarily focus on evading NIDS classifiers, generating NAEs in the NIDS context poses an additional challenge of bypassing advanced adversarial example detectors.
%
{\em To the best of our knowledge, \ND~is the first approach to systematically identify perturbable features and integrate diffusion models into adversarial attacks specifically targeting the NIDS domain.}


The main \textbf{contributions} of this paper are the following:
\begin{itemize}
    \item We develop a novel algorithm for systematically determining perturbable features that can be manipulated while preserving the coherence and structural integrity of network traffic flow data, enabling the generation of effective adversarial examples.
    \item We propose a novel approach for generating Natural Adversarial Examples (NAEs) in the NIDS context. Our approach exploits a diffusion model to produce realistic and imperceptible adversarial network flow instances.
    \item We demonstrate that a DL-based NIDS can be effectively deceived by comparing \ND~with the state-of-the-art conventional attack strategies such as Fast Gradient Sign Method (FGSM), Projected Gradient Descent (PGD), and Auto Conjugate Gradient (ACG), by using three benchmark NIDS datasets and various DL model architectures. 
    Our experimental results demonstrate that \ND~achieves up to a 29.93\% higher attack success rate compared to the baseline attacks across all benchmark NIDS datasets.
    
    \item We evaluate the effectiveness of \ND~in evading the state-of-the-art adversarial example (AE) detectors.
    Experimental results reveal that our proposed method significantly reduces AE detection performance by up to 0.267 in AUC-ROC compared to baseline adversarial attacks on the MANDA detector and 0.534 on the Artifact detector.
    
\end{itemize}

The remainder of the paper is organized as follows: Section~\ref{sec:relatedwork} provides an overview of the background.
Section~\ref{sec:methodology}, details our methodology for designing \ND. Section~\ref{sec:evaluation} presents the evaluation of \ND. Section~\ref{sec:conclusion} summarizes our findings and discusses potential directions for future research.


%% file: related-work.tex
\section{Background and Related Work} \label{sec:relatedwork}
\subsection{DL-based Network Intrusion Detection}
Deep Neural Networks (DNNs) are increasingly employed in NIDS due to their ability to reduce feature engineering efforts while achieving high detection accuracy~\cite{ferrag2020deep}.
Niyaz \textit{et al.}~\cite{javaid2016deep} employed sparse autoencoders for self-taught feature extraction in traffic flows, achieving an F1-score of 0.98 on the NSL-KDD dataset~\cite{tavallaee2009detailed}.
Faker and Dogdu~\cite{faker2019intrusion} developed a Multi-Layer Perceptron (MLP) for intrusion detection on the CICIDS2017 dataset~\cite{sharafaldin2018toward}. Despite its simplicity, their MLP outperformed traditional models such as Random Forest and SVM, achieving notably higher detection rates.

Vinayakumar \textit{et al.}~\cite{vinayakumar2019deep} similarly showed that MLP-based NIDS outperformed a range of traditional machine learning approaches, highlighting the effectiveness of deep learning for intrusion detection.
Naseer \textit{et al.}~\cite{naseer2018enhanced} developed IDS using Convolutional Neural Networks (CNNs), autoencoders, and Long Short-Term Memory Networks (LSTMs), achieving test accuracies of 85\% and 89\% with CNN and LSTM models on the NSL-KDD dataset, demonstrating DNNs effectiveness for NIDS.
Sun \textit{et al.}~\cite{sun2020dl} proposed DL-IDS, a hybrid CNN-LSTM architecture that captures spatial and temporal features of network traffic, achieving 98.67\% accuracy on the CICIDS2017 dataset.

Traditional industrial IDS solutions, such as  Snort~\cite{snort10.5555/1039834.1039864}, Suricata~\cite{suricata}, and Zeek~\cite{zeek} primarily rely on rule- or signature-based detection, which limits their ability to identify novel attacks. In contrast, deep learning-based IDSs have gained traction for learning complex patterns and detecting unseen threats. In this work, we focus on DL-based IDS models because they are more adaptable to evolving threats and are increasingly being studied for deployment as robust NIDS.
\subsection{Adversarial Examples}
\label{lbl:AE}
The term ``adversarial example'' (AE) was introduced by Szegedy \textit{et al.}~\cite{szegedy2013intriguing}, who showed that small, imperceptible perturbations to inputs can mislead classifiers into incorrect predictions. 
AEs are deliberately crafted inputs intended to deceive target models into producing erroneous outputs.
The adversary's objective is to create these examples by introducing minimal perturbations to the input data attributes or features. Given $f(x) = y$, where $f$ is a model that
maps a non-adversarial input sample $x \in \mathbb{R}^n$ to its correct class/label, $y$, and $n$ is the number of features, an adversary tries to find the minimum perturbation $\Delta x$ to cause the output of $f(x + \Delta x)$ to be a specific label, $y_{adv} \ne y$. 
Therefore, an adversarial example $x_{adv} = x + \Delta x$ can be crafted by solving the following optimization problem:
\begin{equation}
\label{eq:adv}
\begin{split}
\min_{\Delta x} c \Delta x + J(f(x + \Delta x), y_{adv}) \\
\text{such that } (x+\Delta x) \in [0,1]^n
\end{split}
\end{equation}
where $c$ is a constant and $J$ is the loss function. Adversarial examples are commonly studied on unconstrained domains such as computer vision~\cite{carlini2017towards,papernot2016distillation}, specifically in image classification applications, where a constraint in finding adversarial examples is that the added noise should be imperceptible to human eyes.
However, most domains are subject to constraints. In this work, we generate adversarial examples in the network domain 
for intrusion detection applications by identifying and enforcing the domain-specific constraints. 

\subsection{Diffusion Models}
\label{lbl:DM}
Denoising Diffusion Probabilistic Model (DDPM)~\cite{ho2020denoising} is a type of generative model that utilizes diffusion processes, inspired by principles in physics, to learn complex data distributions and generate high-quality samples~\cite{sohl2015deep}.  
This model type relies on two Markov processes: a \textit{forward} process and a \textit{reverse} process.
In the \textit{forward} process, Gaussian noise is incrementally added to the data over $T$ steps, 
gradually eliminating its structure and transitioning from the complex data distribution to a Gaussian distribution. 
In the \textit{reverse} process, each step undoes the corresponding forward step by removing the diffusion noise to reconstruct the original data distribution. Since the reverse process is mathematically intractable, it is approximated using DNNs. 
Diffusion models have a wide range of applications across various domains. For instance, Liu \textit{et al.}~\cite{liu2023more} used it for image and text-based guidance. Stable diffusion~\cite{rombach2022high}
produced professional-quality artistic paintings with user-specified text prompts. 
DiffPure~\cite{nie2022DiffPure}  leveraged DDPM for adversarial purification. It takes an adversarial example as input and reconstructs the clean image using a reverse generative process.
Recent work, such as TabDDPM~\cite{TabDDPM10.5555/3618408.3619133}, have demonstrated the effectiveness of diffusion models in generating high-quality synthetic tabular data.
AdvDiffuser~\cite{Chen_2023_ICCV}
employed diffusion models to 
generate Natural Adversarial Examples (NAEs)~\cite{NAE-DBLP:conf/iclr/ZhaoDS18} that are semantically more meaningful than conventional AEs and lie on the original data manifold. 

\subsection{Adversarial Attacks and Defenses on NIDS}
\label{lbl:adv-attack}

Early adversarial example generation methods, such as the linear search by Szegedy \textit{et al.}~\cite{szegedy2013intriguing}, were computationally expensive and impractical.
To address this, Goodfellow \textit{et al.} proposed the Fast Gradient Sign Method (FGSM)~\cite{goodfellow2014explaining}, which generates adversarial examples via a single-step perturbation in the direction of the gradient sign.
The Jacobian-based Saliency Map Attack (JSMA), proposed by Nicolas Papernot \textit{et al.}~\cite{papernot2016distillation}, is an iterative, targeted algorithm that uses a saliency map to identify minimal features with the greatest impact on classification. Carlini and Wagner~\cite{carlini2017towards} (CW) introduced an optimization-based algorithm for generating adversarial examples across various norms using different objective functions.
Basic Iterative Method (BIM)~\cite{kurakin2018adversarial} and Projected Gradient Descent (PGD)~\cite{PGDmadry2018towards} are improvements of FGSM where the algorithm iteratively
increases the amount of perturbation to cause misclassification.
Yamamura~\textit{et al.}~\cite{ACG-DBLP:conf/icml/YamamuraSTHMOIF22} introduced the Auto Conjugate Gradient (ACG) attack, which uses the conjugate gradient method to handle ill-conditioned problems that reduce the effectiveness of traditional steepest descent–based attacks for generating adversarial examples.

Most existing methods~\cite{goodfellow2014explaining,kurakin2018adversarial,PGDmadry2018towards} that generate adversarial examples focus on compromising models for image-based applications. \textit{In contrast, research on evading deep learning-based NIDS remains limited}.
Wang \textit{et al.}~\cite{wang2018deep} demonstrated that four white-box attacks could effectively bypass MLP-based intrusion detectors trained on the NSL-KDD dataset~\cite{tavallaee2009detailed}, highlighting their vulnerability to transferable adversarial examples.
Ibitoye \textit{et al.}~\cite{ibitoye2019analyzing} observed the effect of three adversarial attacks (FGSM, BIM, and PGD) on their deep learning-based IDS for IoT networks.
Teuffenbach \textit{et al.}~\cite{teuffenbach2020subverting} improved adversarial crafting algorithms by integrating domain-specific constraints. Using iterative FGSM and CW attacks on the CICIDS2017 dataset, they demonstrated successful evasion of DL-based NIDS.
Sheatsley \textit{et al.}~\cite{sheatsley2022adversarial}
modified the JSMA~\cite{papernot2016distillation} algorithm and proposed a method to sketch adversarial perturbations on network features. Their results revealed that network constraints do not significantly improve robustness against adversarial attacks.

Given the prevalence of adversarial attacks and their potential impact, researchers have developed numerous defense mechanisms to detect and mitigate adversarial examples.
MANDA~\cite{Manda9709532} uses manifold learning to classify benign and malicious data, flagging mismatches as adversarial. However, it focuses on feature-level attacks with domain constraints and assumes the defender's access to malicious data, which is not always practical.
Artifact, proposed by Feinman \textit{et al.}~\cite{feinman2017detecting}, is a state-of-the-art adversarial example detection method. Unlike MANDA, it employs kernel density estimation and Bayesian neural network uncertainty as criteria for detecting adversarial examples.
Grosse \textit{et al.}~\cite{DBLP:journals/jmlr/GrettonBRSS12} employed the Maximum Mean Discrepancy (MMD) test, a hypothesis test used to assess whether two data sets originate from the same underlying distribution to detect adversarial examples. Their motivation is that adversarial and clean examples belong to different distributions due to the adversarial perturbations.

\subsection{Synthetic Network Traffic Evaluation}
\label{lbl:quality_evaluation_of_generated_data}
In this section, we summarize the key criteria and evaluation methods for assessing the quality and accuracy of the generated network traffic data. 
To benchmark \ND's adversarial outputs, we leverage prior surveys to identify essential metrics for evaluating traffic quality.

Several prior works have noted that \textit{there is no standard method to evaluate the quality of generated network traffic,} 
regardless of the generation source: emulator, hardware, or generative AI models~\cite{DBLP:journals/csur/AdelekeBG23,agrawal2024review,DBLP:journals/csur/HalvorsenICG24,schoen2022towards}. 
Oluwamayowa \textit{et al.}~\cite{DBLP:journals/csur/AdelekeBG23} 
highlighted that software and hardware-based traffic generation tools often lack formal evaluation methodologies.
Agrawal \textit{et al.}~\cite{agrawal2024review} reviewed Generative Adversarial Network (GAN)-based flow-level traffic generation and noted that existing studies employ diverse evaluation methods, including pattern matching, histogram distance, and domain-specific sanity checks.

Halvorsen \textit{et al.}~\cite{DBLP:journals/csur/HalvorsenICG24} 
identified a similar gap in evaluating data generated by models such as GANs, diffusion models, and variational auto-encoders (VAEs) for IDS. Based on their review of 27 studies, two primary evaluation strategies are commonly used: (i) training a classifier on synthetic data and testing on real data, or vice versa, to assess the realism of generated samples; and (ii) applying statistical metrics to quantify distributional differences between real and synthetic datasets.
Schoen \textit{et al.}~\cite{schoen2022towards} emphasized the importance of multi-view evaluation for assessing synthetic traffic quality. They proposed three key criteria: (i) \textit{Realism (Fidelity):} synthetic samples should reflect the distribution of real data; (ii) \textit{Diversity (Fairness):} generated data should capture the variability of the original dataset; and (iii) \textit{Originality (Authenticity):} synthetic samples should be distinct from real instances. 
We chose these last three metrics. Our rationale for adopting them and their application for \ND's evaluation is detailed
in Section~\ref{subsubsection:eval_adv_data}.




%% file: modelsandassumptions.tex
\section{Models and Assumptions} \label{sec:m&a}
In this section, we first introduce the targeted NIDS model, followed by a detailed adversarial threat model.
\subsection{System Model} \label{lbl:system_model}
\begin{figure}[htbp]
  \centering
  \includegraphics[width=0.48\textwidth]{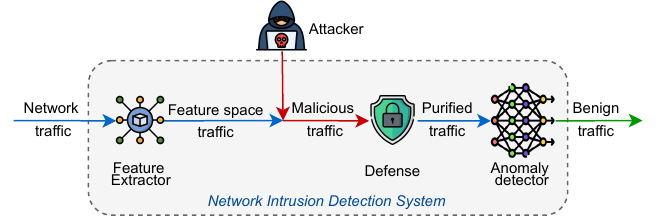}
  \caption{Overview of a DL-based NIDS.}
  \label{fig:DL-NIDS}
\end{figure}

We consider a NIDS (the targeted victim) to be a deep learning (DL) model that detects network anomalies via traffic analysis.  A typical architecture of a DL-based NIDS is illustrated in Figure~\ref{fig:DL-NIDS}. Usually, a NIDS operates as a passive component, gathering information from network flows in the network of interest, minimally interfering with the network traffic it monitors. A DL-based NIDS mainly comprises the following key modules: 




\begin{itemize}
    \item \textit{Flow Collector:} This module captures the ongoing raw network traffic (\textit{i.e.} packets of a communication system) and maps them to different network flows. In our work, all the raw network traffic is stored as PCAP files. 
    \item \textit{Feature Extractor:} This module converts network flows into feature vectors by calculating statistical metrics that implicitly represent the state of the flows. CICFlowMeter~\cite{lashkari2017characterization} is a feature extraction tool.
    \item \textit{Anomaly Detector:} 
    This module consists of a DL-based binary NIDS model trained with both benign and malicious network traffic flow instances in a training phase. The runtime traffic instances are fed into this model, which predicts one for malicious traffic, otherwise zero. Either an alert will be generated to notify the system administrators, or a defense mechanism will be triggered if an input instance is classified as malicious by the IDS model.
\end{itemize}

\subsection{Threat Model}
\label{lbl:threat_model}

We focus on scenarios where an adversary tries to
compromise the DNN model adopted by the NIDS, making it perform misclassification and evade
detection of their malicious traffic. For instance, the adversary aims to cause the NIDS to either
classify a malicious network traffic instance as benign (\textit{i.e.}, a False
Negative) or classify a benign one as malicious (\textit{i.e.}, a False
Positive). 
It is important to note that \ND~operates at the flow level. In our implementation, each network flow is represented by a set of features extracted from raw network traffic data. 
If malicious traffic bypasses the detection mechanism, it can inflict other types of attacks (\textit{e.g.}, SQL injection, cross-site scripting) on the protected system. On the other hand, if NIDS predicts benign traffic as anomalous/malicious and blocks it, this may lead to severe denial-of-service attacks on the victim system.
The adversary can achieve this by slightly adjusting the features extracted from a traffic flow, $x$, which is the input to the NIDS model, and creating an adversarial example $x_{adv}$ using Equation~\ref{eq:adv} as discussed in Section~\ref{lbl:AE}.
For instance, the attacker can apply subtle perturbations to the mean forward inter-arrival
time feature such that the perturbed feature preserves the statistical properties.

Depending on the adversary's prior knowledge about the target DL-based system, attacks can be categorized into three types:
(i) \textit{white-box attacks} where the adversary has full knowledge of the target model’s architecture and trained parameters, enabling direct gradient-based optimization to craft adversarial perturbations; 
(ii) \textit{gray-box attacks} where attackers know the model’s architecture but lack access to its parameters or training data, requiring them to train surrogate models and exploit the transferability of adversarial examples; 
and (iii) \textit{black-box attacks} where adversaries do not have any knowledge about the architecture, parameters, or training data of the target DL model. Instead, adversaries rely on querying the model to observe its outputs to iteratively refine AEs. 

Among the three attack types, white-box attacks are the most potent, as they enable the generation of highly effective and stealthy adversarial examples by leveraging full access to the target model.
Although comparatively difficult to orchestrate, given the widespread availability of open-source models and the ability of adversaries to build closely mimicking surrogate models, this remains a highly feasible attack vector.
Therefore, in this paper, we adopt white-box attacks as the threat to challenge and defeat the NIDS. 
While we focus on white-box attacks to demonstrate \ND’s capability, it can also extend to gray-box (using surrogate model gradients) and black-box settings (using query-based gradient estimation) without altering the rest of the framework. 


%% file: methodology.tex
\section{NetDiffuser Framework} \label{sec:methodology}
\subsection{Overview of \ND}
\label{lbl:attack-frame}
Unlike traditional adversarial attacks that aim to cause misclassifications with minimal perturbation, \ND~focuses on generating Natural Adversarial Examples (NAEs). 
These are adversarial inputs that remain close to the real data distribution while still deceiving the anomaly detector model. We denote \( x \) as the original benign input, \( x_{\text{adv}} \) as a traditional adversarial example generated using gradient-based attacks, and \( x_{\text{nae}} \) as the NAE crafted using \ND. The core idea is that \( x_{\text{nae}} \) not only fools the classifier but also preserves the semantics and statistical properties of real network traffic, making it harder to detect by the adversarial detectors.
\begin{figure}[htbp]
\centerline{\includegraphics[width=0.5\textwidth]{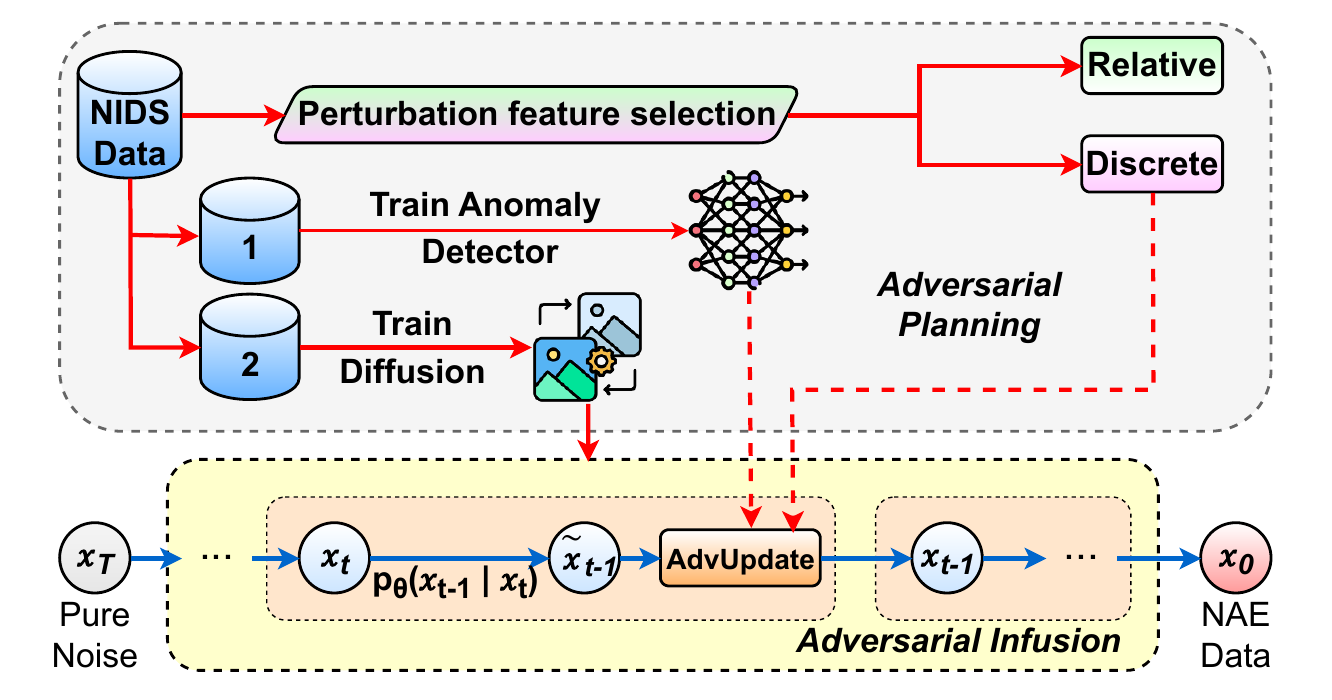}}
\caption{The schematic of \ND.} 
\label{fig:methodology}
\end{figure}
We illustrate a high-level overview of our proposed attack framework, \ND~in Figure~\ref{fig:methodology}.
The framework is structured into two primary stages: \textit{Adversarial Planning}~(\ref{sec:adv-plan}) and \textit{Adversarial Infusion}~(\ref{sec:reverse_diffusion}).
The first stage, \textit{Adversarial Planning}, involves analyzing the NIDS dataset to categorize features and identify those suitable for perturbation within the network domain, as well as training the target anomaly detector and diffusion models,
which form the foundation for crafting adversarial examples.
The second stage, \textit{Adversarial Infusion}, utilizes the trained models and selected features to generate \( x_{\text{nae}} \) through a diffusion-based process that integrates perturbations at each denoising step.

\subsection{Adversarial Planning}
\label{sec:adv-plan}
The \textit{Adversarial Planning} stage encompasses two critical tasks for generating natural adversarial examples: selecting perturbation features and training models.

\textbf{Perturbation Feature Selection:} Unlike adversarial attacks on image classifiers, adversarial samples targeting NIDS must adhere to specific domain constraints~\cite{ConstDomain:journals/corr/abs-2011-01183} to preserve the functionality and integrity of the samples. This implies that (i) only a limited subset of features can be perturbed, and (ii) the altered features must remain consistent with the inherent properties of the original samples. Therefore, to enforce network constraints, we 
focus on identifying key features to craft adversarial examples that closely resemble legitimate traffic flows. We begin this process by partitioning diverse features representing network traffic flow characteristics into 
two categories. 

(1) \PF~features: 
    These features are relatively 
    independent of all other features for all network traffic data despite of the feature extractors that are utilized. 
    %
    For example, features such as \textit{Fwd Packet Length Min} and \textit{Bwd Packet Length Min} extracted from raw traffic data using CICFlowMeter representing the minimum packet lengths for forward and backward traffic respectively, 
    are independent of flow-level metrics, such as \textit{Packet Length Mean} or \textit{Packet Length Std}. 
(2) \SF~features: These features exhibit 
stronger dependencies 
    on other features that are fundamental to the natural structure and inherent relationships within the data.
    For example, \textit{Flow IAT Mean}, \textit{Flow IAT Std}, and \textit{Flow IAT Max} are 
    strongly dependent because the mean, standard deviation, and maximum are all calculated from the same packet intervals in the flow. 

Our process of partitioning the features is described in Algorithm~\ref{alg:feature_categorization}. It operates by systematically analyzing the relationships between features through observations of the data to identify and categorize them into \PF~and \SF~feature groups. 
It is important to note that \PF~features exhibit lower 
dependency compared to those in the \SF~group. 
For instance, while modifying a feature such as \textit{Fwd Packet Length Min} can influence aggregated metrics like \textit{Packet Length Mean} or \textit{Packet Length Std}, our analysis of pairwise correlations reveals that the dependence between these discrete features and the aggregated metrics is significantly weaker than the interdependencies observed within the \SF~group. 
This relative independence is sufficient for our perturbation strategy, as it helps maintain overall data integrity while allowing effective adversarial manipulation. 

\begin{algorithm}
\begin{algorithmic}
\small
\State \textbf{Input:} Dataset, $\mathcal{D} \in \mathbb{R}^{m\times n}$ and set, $U = \{u_1,u_2, \cdots, u_n\}$ of all feature identifiers.
\State \textbf{Output:} \textit{Discrete} Features, $\mathcal{P}$; \textit{Relative} Features, $\mathcal{S}$
\end{algorithmic}

\begin{algorithmic}[1]
\small
    \State \text{Construct } $\mathcal{D}' $
    \label{algo1:step-X}
   
    \State $r_{ij} \gets \frac{\text{Cov}(v_i, v_j)}{\sigma_{v_i} \sigma_{v_j}}, \quad \forall i, j \in \{1, 2, \dots, n\} \text{ and } v_i,v_j \in \mathcal{D}'$ \label{algo1:step-r}
    
    \State $D \gets [d_{ij}] \in \mathbb{R}^{n \times n}$, where
    $d_{ij}=\sqrt{2(1 - r_{ij})}; \quad \forall i, j \in \{1, 2, \dots, n\}$ \label{algo1:step-d}
    
    \State $\text{Dendogram}, \mathcal{T} = \text{AgglomerativeClustering}(D)$ \label{algo1:step-T} 
    

    \State $h^* \gets \underset{h}{\arg\max} \Big( I^h_{\text{CH}} - I^{h'}_{\text{CH}} \Big),
    \text{where } h,h' \in [h_{\text{min}}, h_{\text{max}}]$ $\text{ are heights at } \mathcal{T}$ \label{algo1:step-h}
    
    \State $\mathcal{P} \gets \{u_i \mid u_i \in \text{Clusters at } h^* \text{ of } \mathcal{T}\}$ \label{algo1:step-P} 
    
     \State $\mathcal{S} \gets \{u_i \mid u_i \notin \text{Clusters at } h^* \text{ of } \mathcal{T}\}$ \label{algo1:step-S}
    
    \State \Return $\mathcal{P}, \mathcal{S}$
\caption{Feature Categorization Algorithm}
\label{alg:feature_categorization}
\end{algorithmic}
\end{algorithm}

Algorithm~\ref{alg:feature_categorization} takes as input the data matrix $\mathcal{D} \in \mathbb{R}^{m\times n}$, where $m$ is the number of network traffic flows and $n$ is the number of features of each flow, and set $U = \{u_1, u_2, \cdots u_n\}$ where $u_j$ is the $j$-th feature identifier. Step~\ref{algo1:step-X} computes the transpose of $\mathcal{D}$, denoted as $\mathcal{D}'$. 

   
Intuitively, \SF~features could be 
    identified by clustering the features that are 
    more linearly correlated with others.
    Step~\ref{algo1:step-r} computes the
    Pearson Correlation Coefficient (PCC), $r_{ij}$ for all pairs of feature vectors, $v_i, v_j \!\in\! \mathcal{D}'$ where $\forall\!i,j\in\!\{1,2,\cdots n \}$, capturing the degree of linear dependency between them. 
    The use of PCC is motivated by the experimental methodology employed in~\cite{ConstDomain:journals/corr/abs-2011-01183} to identify correlations among features. 
    Step~\ref{algo1:step-d} constructs a distance matrix, $D$ representing pairwise distances between features calculated using $d_{ij}$= $\sqrt{2(1 - r_{ij})}$.
    Step~\ref{algo1:step-T} employs the Agglomerative Hierarchical Clustering~\cite{davidson2005agglomerative} technique to iteratively group features based on $D$. 
    We chose this clustering approach as it does not require a predetermined number of clusters. Instead, it generates a hierarchical tree structure called \textit{dendrogram} that visually represents the relationships between features. 
    We denote the dendrogram we obtained from Step~\ref{algo1:step-T} as $\mathcal{T}$.

    To determine the optimal number of clusters, we utilize the Calinski-Harabasz (CH) Index~\cite{calinski1974dendrite}, denoted as $I_{CH}$, which evaluates cluster quality by measuring the ratio of between-cluster variance to within-cluster variance. This ensures that the clusters are both well-separated and internally cohesive.
    In Step~\ref{algo1:step-h} of Algorithm~\ref{alg:feature_categorization}, we fetch two different sets of clusters by cutting the dendrogram 
    at two heights, $h$ and $h'$ and compute their CH Indices, $I^h_{CH}$ and $I^{h'}_{CH}$ respectively. 
    We choose the height $h$ that maximizes the difference between $I^h_{CH}$ and $I^{h'}_{CH}$ indices. We denote this optimal height as $h^*$. We hypothesize that selecting the height $h$ that maximizes the difference between the CH indices $I^h_{CH}$ and $I^{h'}_{CH}$ ensures the most meaningful partitioning of features in the dendrogram by optimizing the trade-off between intra-cluster compactness and inter-cluster separation. 
    The dendrogram, $\mathcal{T}$ 
    is then cut at the height (\(h^*\)), yielding a set of clustered and non-clustered features.
    Notably, while previous works rely on manually selecting perturbation features, an approach that is often subjective and may not generalize well across diverse NIDS datasets, our systematic, data-adaptive method consistently yields robust feature groupings. In Steps~\ref{algo1:step-P} and~\ref{algo1:step-S}, we organize the features into two sets: $\mathcal{P}$, containing all features within clusters as \SF~features, and $\mathcal{S}$, containing all features outside clusters as \PF~features.

\textbf{Model Training:}
The next step is to train a network traffic Anomaly Detection Model 
and a diffusion model. 
The NIDS dataset $\mathcal{D}$ is partitioned into separate subsets: $\mathcal{D}_1$ and $\mathcal{D}_2$, to independently train an anomaly detector and a diffusion model. Both $\mathcal{D}_1$ and $\mathcal{D}_2$ have 50\% of data in $\mathcal{D}$. 

\noindent \textit{Anomaly Detection Model \((f)\):} This model is trained using  $\mathcal{D}_1$ to classify network traffic flow as either normal or anomalous, establishing a baseline for the anomaly detection task. In later stages, it serves as the target model for crafting adversarial examples.

\noindent \textit{Diffusion Model \((\mathcal{M}_\theta)\):} The diffusion model is trained to learn the distribution of the data, $\mathcal{D}_2$. During the training phase, the model, $\mathcal{M}_\theta$ undergoes a forward diffusion process, in which clean data sample, $x_{0} \in \mathcal{D}_2$ is gradually transformed into noisy Gaussian representations ($x_{T}$) over a series of $T_{fwd}$ steps. 
    The trained diffusion model is later employed in the Adversarial Infusion stage, as detailed in Section~\ref{sec:reverse_diffusion}, to generate adversarial examples. These examples are subsequently used to challenge the adversarial example detector before being processed by the anomaly detection model.
    


\subsection{Adversarial Infusion} \label{sec:reverse_diffusion}

The \textit{Adversarial Infusion} stage focuses on crafting adversarial examples by leveraging the trained diffusion model, $\mathcal{M}_\theta$, and utilizing the \textsc{AdvUpdate} method to iteratively inject perturbations into the \PF~features that were identified using Algorithm~\ref{alg:feature_categorization} as described in Section~\ref{sec:adv-plan}. 
The objective is to generate traffic flows that can effectively deceive the network anomaly detection model by maintaining the semantic integrity of the original flows. This stage ensures that the perturbations introduced to the subset of features by the \textsc{AdvUpdate} method undergo a refinement process during the denoising steps of the diffusion model, resulting in adversarial example flows that closely resemble natural network traffic.

This stage begins with the noisy representation of 
a flow, $x_{T}$, generated at the final step of the forward diffusion process during model training of $\mathcal{M_\theta}$ as outlined in Section~\ref{sec:adv-plan}.
The trained model, $\mathcal{M_\theta}$, generates a partially denoised version of the 
noisy representation at each reverse time step. 
To inject adversarial intent, adversarial perturbations are introduced to each denoised sample. 
This step serves as a modular update function that supports multiple attack methods, including FGSM, PGD, and ACG, to iteratively introduce perturbations to the denoised sample.
\textit{We posit that this update function is compatible with other gray-box or black-box attack methods, thereby demonstrating the modularity and extensibility of our proposed framework}

These perturbations are calculated to maximize the adversarial loss of the target anomaly detection model, $f$. We restrict the application of the perturbations only to the set of \PF~features determined by  
Algorithm~\ref{alg:feature_categorization}. 
Features outside this designated subset, which might disrupt the semantics of the network flow 
and increase the likelihood of detection by adversarial example (AE) detectors, 
are left unaltered during this process. This approach is consistent with research~\cite{su2019one}, highlighting that adversarial examples can be effectively generated by modifying only a minimal subset of input features.

Once the perturbations are infused into the \PF~features, the modified sample undergoes the reverse diffusion step of the trained diffusion model, $\mathcal{M}_\theta$. This denoising process is repeated iteratively to gradually refine the sample, ensuring it closely resembles a realistic representation of the original data.
Thus, the final generated data point, $x_{0}$, maintains the overall structure and coherence of legitimate data while having adversarial properties. 

\begin{algorithm}[ht]
    \caption{The overall algorithm of \ND}
    \label{alg:net_diffuser}
    \small
        \begin{algorithmic}[0]
        \State \textbf{Input:} \\
        \quad $\mathcal{D}$: 
        Dataset for feature selection and training; \\
        \quad $f$: Target anomaly detector; \\ 
        \quad $T_{\text{fwd}}$: Total number of forward diffusion steps; \\
        \quad $T_{\text{rev}}$: Total number of reverse diffusion steps; \\
        \quad $attack$: Adversarial method for perturbation; \\
        \quad $\epsilon$: Maximum allowed perturbation magnitude; \\
        \quad $\alpha$: Perturbation step size for adversarial updates
        \State \textbf{Output:}
        \quad Adversarial example, $x_{\text{adv}}$
        \end{algorithmic}
        \begin{algorithmic}[1]
            \State $\mathcal{P}, \mathcal{S} \gets \text{categorize features of } \mathcal{D} \text{ using Algorithm~\ref{alg:feature_categorization}}$\label{algo2:feature-cat}

            \State Initialize $x_0 \gets \text{sample from dataset}$\label{algo2:sampling}

            \State \textbf{for} $t \in [1, 2, \dots, T_{fwd}]$\label{algo2:fwd-loop-st} \textbf{do}
                \State \quad Sample $\epsilon \sim \mathcal{N}(0, I)$
                \State \quad $x_t \gets \sqrt{1 - \beta_t} \cdot x_{t-1} + \sqrt{\beta_t} \cdot \epsilon$
            \State \textbf{end for}\label{algo2:fwd-loop-end}

            \State $x_{T_{rev}} \gets x_t$\label{algo2:xt}
                \State \textbf{for} $t \in [T_{rev}, T_{rev}-1, \dots, 1]$ \textbf{do}\label{algo2:rev-loop-st}
                
                    \State \quad 
                    $\tilde{x}_{t-1} \gets p_\theta(x_{t-1} \mid x_t, \mathcal{M}_\theta)$
                    \label{algo2:rev-pred}
                    \State \quad \textbf{for} $i \in [0, 1, \dots, I-1]$ \textbf{do}\label{algo2:pgd-loop-st}
                        \State \quad \quad \textbf{if} $\underset{i}\arg \max f(\tilde{x}^{(i)}_{t-1}) \neq y$ \textbf{then}
                        \label{algo2:f-pred}
                        \State \quad \quad \quad \textbf{break}
                        \State \quad \quad \textbf{end if}
                        \label{algo2:f-pred-end}


                         \State \quad \quad $\tilde{x}^{(i+1)}_{t-1} \gets \textsc{AdvUpdate}\big(\tilde{x}^{(i)}_{t-1}), f, y, \epsilon, \alpha, \mathcal{S}, attack\big)$ \label{algo2:adv-update}


                    \State \quad \textbf{end for}\label{algo2:pgd-loop-end}
                    \State \quad $x_{t-1} \gets \tilde{x}_{t-1}$\label{algo2:rev-restart}
                \State \textbf{end for}
                \label{algo2:rev-loop-end}
            \State $x_{\text{nae}} \gets x_0$\label{algo2:x-adv}
            \State \Return $x_{\text{nae}}$
    \end{algorithmic}
\end{algorithm}
\subsection{The \ND~Algorithm}
\label{sec:netdiffuse}
Inspired by the work in~\cite{Chen_2023_ICCV}, we adapt the AdvDiffuser algorithm as the foundation for our proposed \ND~algorithm.
We outline the complete algorithm of our \ND~attack in Algorithm~\ref{alg:net_diffuser}. The algorithm accepts as input the NIDS dataset, $\mathcal{D}$, the target network anomaly detector, $f$, and parameters such as the total number of forward ($T_{fwd}$) and reverse ($T_{rev}$) diffusion steps, the number of adversarial refinement iterations,
$I$, 
perturbation step size $\alpha$ and maximum perturbation magnitude $\epsilon$.

The \textit{Adversarial Planning} stage encompasses Steps~\ref{algo2:feature-cat} to~\ref{algo2:fwd-loop-end} in Algorithm~\ref{alg:net_diffuser}.
Step~\ref{algo2:feature-cat} categorizes all features from the dataset using Algorithm~\ref{alg:feature_categorization} to extract the \PF~features. Step~\ref{algo2:sampling} initializes $x_0$ with a random sample from the original flow. 
In Steps~\ref{algo2:fwd-loop-st} to~\ref{algo2:fwd-loop-end}, the sample $x_0$ undergoes $T_{fwd}$ number of forward diffusion steps, where noise sampled from a Gaussian distribution, $\mathcal{N}(0, I)$, is progressively added. This transforms $x_0$ into an isotropic Gaussian representation, which serves as the starting point for the Adversarial Infusion stage.

The \textit{Adversarial Infusion} stage comprises Steps~\ref{algo2:xt} to~\ref{algo2:x-adv}. This stage begins by initializing $x_{T_{rev}}$ with the Gaussian representation, $x_t$ produced at the end of 
{\em forward process of the diffusion model}.
At each reverse diffusion step $t$, an intermediate denoised sample, $\tilde{x}_{t-1}$ of the noisy representation is predicted using the trained diffusion model, \(\mathcal{M}_\theta\) in Step~\ref{algo2:rev-pred}. 
Steps~\ref{algo2:pgd-loop-st} to~\ref{algo2:pgd-loop-end} iteratively apply perturbations using the PGD method. In Step~\ref{algo2:f-pred}, the PGD method terminates upon verifying if the $i$-th iteration produces the maximum prediction error for the anomaly detector, $f$. 
In Step~\ref{algo2:adv-update}, adversarial perturbations are computed using a modular update function, \textsc{AdvUpdate}, which supports different adversarial attack strategies such as FGSM, PGD, and ACG. This function dynamically computes the loss of the anomaly detector \(f\) with respect to the true label \(y\) and its gradient relative to the current intermediate sample \(\tilde{x}_{t-1}\) to derive the perturbation \(\tilde\Delta_{t-1}^{(i)}\). These perturbations are applied selectively to the \PF~features \(\mathcal{S}\), as determined by our feature categorization algorithm. 
After that, $x_{t-1}$ is fed into the reverse diffusion step. This process is repeated for all reverse diffusion steps $t \in [T_{rev}, T_{rev}-1, \dots, 1]$ to produce $x_0$. Thus,  the initial noisy representation of the sample, $x_t$, is transformed into a realistic adversarial example, $x_{nae}$, in Step~\ref{algo2:x-adv}.
These steps can be followed iteratively to generate large number of adversarial examples by an adversary to inject into the network of interest to attack the NIDS.  

%% file: evaluation.tex
\section{Evaluation}
\label{sec:evaluation}
This section evaluates the performance of our \ND~attack. Section~\ref{sec:experiment} provides the detailed experimental setting.
Section~\ref{sec:defenses} describes the state-of-the-art defenses (\textit{i.e.,} adversarial example detectors) used to assess the effectiveness of our \ND~attack. Section~\ref{sec:results} analyzes the performance 
of the adversarial attacks, compares the attack efficacy in the presence of adversarial detectors, and evaluates the quality of adversarial examples generated by \ND. 
\subsection{Experiments}
\label{sec:experiment}


\subsubsection{Datasets}



We selected three widely used datasets for intrusion detection systems and network traffic analysis: CICIDS2017~\cite{sharafaldin2018toward}, CICDDOS2019~\cite{sharafaldin2019developing}, and UNSW-NB15~\cite{7348942}. 
In this study, we frame the task as a binary classification problem, categorizing network traffic as either normal (benign) or malicious.

{\bf \textit{CICIDS2017}}: It is a widely recognized benchmark dataset used for NIDS.
It contains over 2.8 million network flow records with more than 80 attributes extracted from real-world traffic. We preprocessed the data by selecting a balanced subset of 60,000 instances and removing redundant features to improve model performance, resulting in a dataset with 77 features. 


{\bf \textit{CICDDoS2019:}} 
It includes a wide range of DDoS attack types along with legitimate network traffic. The dataset contains over 80 flow-based features extracted from network traffic, capturing packet-level, time-based, and connection-based attributes. For our experiments, we utilize a balanced subset of 1 million instances, comprising 500,000 benign instances and 500,000 malicious instances. After removing redundant features, the final dataset includes 78 distinct features, making it an ideal resource for developing and benchmarking DDoS detection techniques.

{\bf \textit{UNSW-NB15}}: This dataset includes nine distinct attack categories: Fuzzers, Analysis, Backdoors, DoS, Exploits, Generic attacks, Reconnaissance, Shellcode, and Worms, in addition to benign traffic. It comprises network packets, flows, and content-based attributes. The dataset was created using the IXIA PerfectStorm testbed~\cite{keysight_perfectstorm}, which integrates realistic traffic patterns with synthetic attack scenarios to simulate diverse network conditions. For our experiments, the dataset is divided into a training set with 175,341 instances and a testing set with 82,332 instances, both reduced to 39 features. The training set consists of 119,341 attack instances and 56,000 benign instances, while the testing set includes 45,332 attack instances and 37,000 benign instances.

\textbf{Preprocessing:} 
We removed flow-identifying attributes (e.g., Flow ID, Source IP, Destination IP, Source Port, Destination Port, and Timestamp) from all datasets prior to training or attack generation, as they do not reflect statistical network traffic behavior.
This ensures the models learn from genuine traffic patterns rather than arbitrary identifiers.

\begin{table}[ht]
\caption{Summary of experimental setup including model architectures, attack configurations, and runtime statistics for training and evaluation.} 

\label{tab:runtime_metrics}
\scriptsize
\centering
\begin{tabular}{l|l}
\hline
\multicolumn{2}{l}{\textbf{Diffusion Model Architecture - \# of hidden units of hidden layers}} \\
\hline
UNSW-NB15 MLP-3L & \{256, 512, 256\} \\
CICIDS2017 MLP-2L & \{1024, 1024\} \\
CICDDoS2019 MLP-2L & \{1024, 1024\} \\
\hline
\multicolumn{2}{l}{\textbf{Diffusion Training Time}} \\
\hline
UNSW-NB15 & 18m 3s \\
CICIDS2017 & 26m 48s \\
CICDDoS2019 & 53m 20s \\
\hline
\multicolumn{2}{l}{\textbf{Anomaly Detector (L represents hidden layers)}} \\
\hline
MLP-1L & \{50\}
\\
MLP-5L & \{50, 50, 50, 50, 50\}
\\
CNN-2L & \makecell[l]{Conv2D(32)→MaxPool→Flatten→Dense(50)} \\
CNN-3L & \makecell[l]{Conv2D(32)→MaxPool→Conv2D(32) \\ →MaxPool→Flatten→Dense(50)} \\
Optimizer & Adam \\
Learning Rate & 0.001 \\
Epochs & 30 \\
\hline
\multicolumn{2}{l}{\textbf{Baseline Adversarial Attack - Hyperparameters}} \\
\hline
FGSM  & $\epsilon$ = 0.3\\
PGD & $\epsilon = 0.3$, $\alpha = 0.01$, $I = 40$ \\
ACG & $\epsilon = 0.3$, $\alpha = 0.1$, $I = 100$ \\
\hline
\multicolumn{2}{l}{
    \shortstack[l]{
      \textbf{Baseline Adversarial Attack - Avg. Runtime on processed samples}
    }
} \\
\textit{Attack Targeting MLP-1L} & \quad \textit{FGSM} \,\,\,\quad\quad\quad \textit{PGD} \,\,\,\,\,\,\quad\quad\quad \textit{ACG} \\
\midrule
UNSW-NB15   & \quad 2.14s \,\,\,\,\,\quad\quad\quad 19.42s \,\,\,\,\quad\quad\quad 4.94s \\
CICIDS2017  & \quad 2.10s \,\,\,\,\,\quad\quad\quad 16.07s \,\,\,\,\quad\quad\quad 5.20s \\
CICDDoS2019 & \quad 4.98s \,\,\,\,\,\quad\quad\quad 1m 58s \,\quad\quad\quad 53.35s \\
\hline
\multicolumn{2}{l}{
    \shortstack[l]{
      \textbf{\ND~– Avg. Runtime for Generating 2000 Adversarial Samples
      }
    }
} \\
\textit{Attack Targeting MLP-1L} & \textit{ND-FGSM} \,\quad\quad \textit{ND-PGD} \,\,\quad\quad \textit{ND-ACG} \\
\hline
UNSW-NB15   & \quad 33s \,\,\,\quad\quad\quad\quad 5m 33s \,\quad\quad\quad 5m 40s \\
CICIDS2017  & \quad 31s \,\,\,\quad\quad\quad\quad 3m 21s \,\quad\quad\quad 22m 23s \\
CICDDoS2019 & \quad 29s \,\,\,\quad\quad\quad\quad 4m 49s \,\quad\quad\quad 23m 23s \\


\hline
\multicolumn{2}{l}{\textbf{Defense - Avg. Runtime}} \\
\hline
MANDA & 50m for 2000 samples \\
Artifact & 3m 4s \\

\hline
\end{tabular}
\end{table}

\subsubsection{Experimental Settings}
We performed all the experiments on a system running Ubuntu 23.04 (64-bit) with an Intel 13th Gen Core i9-13900 processor featuring 12 physical cores and 24 logical cores, supporting 32 threads, with a maximum clock speed of 5.6 GHz. The system was equipped with 62 GB of RAM and an NVIDIA RTX A2000 GPU with 12 GB of VRAM. The anomaly detector model was implemented using both Keras and PyTorch frameworks. 
We evaluated adversarial robustness across a diverse set of anomaly detector architectures, including MLP-1L with a single hidden layer, MLP-5L with five hidden layers, and convolutional models (CNN-2L and CNN-3L) with two and three layers, respectively. 
We employed the TabDDPM framework~\cite{TabDDPM10.5555/3618408.3619133} for the diffusion model and implemented the FGSM, PGD, and ACG attacks using the Adversarial Robustness Toolbox (ART)~\cite{DBLP:journals/corr/abs-1807-01069}.
In our experiments, 16, 30, and 27 discrete features were perturbed for the UNSW-NB15, CICIDS2017, and CICDDoS2019 datasets, respectively, based on our feature categorization algorithm described in Section~\ref{sec:adv-plan}.

The experimental configurations and runtime metrics are summarized in Table~\ref{tab:runtime_metrics}. The diffusion model architecture varied across datasets: a 
three-layer MLP with 256, 512 and 256 neurons for UNSW-NB15, a two-layer MLP with 1024 neurons per layer for CICIDS2017, and a 
two-layer MLP with 1024 neurons per layer for CICDDoS2019. 
The models were trained with a learning rate of 0.001 for 80,000 iterations on the CICIDS2017 and CICDDoS2019 datasets. For the UNSW-NB15 dataset, the diffusion model was trained with a learning rate of 0.003 for 200,000 iterations.
The training time for the diffusion models, as presented in Table~\ref{tab:runtime_metrics}, varied across datasets due to differences in model architecture, such as the number of layers and neurons, as well as the number of data instances. 
Table~\ref{tab:runtime_metrics} also presents the model architectures and training configurations of the anomaly detector, the adversarial attack settings, and the reported runtime of the baseline attacks, \ND~for generating samples, and the runtime for the evaluation of the defense mechanisms.



\subsubsection{Baseline Attacks}
We compared \ND~against widely used adversarial attack methods that are commonly employed. Specifically, we considered three gradient-based attacks: FGSM~\cite{goodfellow2014explaining}, PGD~\cite{PGDmadry2018towards}, and ACG~\cite{ACG-DBLP:conf/icml/YamamuraSTHMOIF22}. These attacks exploit the decision boundaries of the anomaly detector model by operating within a fixed perturbation budget ($\epsilon$) to craft adversarial examples.


\subsubsection{Performance Evaluation Metrics} \label{subsubsection:eval_model}

To evaluate our attack on the binary classification task of distinguishing between benign (normal) and anomalous (malicious) traffic, we used the following performance metrics.

    \textbf{Accuracy (acc):} The proportion of correctly predicted instances among the total instances. It is calculated as 
    $Accuracy= \frac{TP + TN}{TP + TN + FP + FN}$,
    where 
    TP (True Positive) refers to the number of correctly predicted positive (\textit{i.e.}, anomalous) instances; TN (True Negative) refers to the number of correctly predicted negative (\textit{i.e.}, benign) instances; FP (False Positive) refers to the number of benign traffic instances incorrectly predicted as anomalous; and FN (False Negative) refers to the number of anomalous instances incorrectly predicted as benign.

    \textbf{Attack Success Rate (ASR):} In general, ASR measures the relative drop in accuracy due to an adversarial attack. In~\cite{Manda9709532}, it is calculated as: 
    $\text{ASR} = \frac{(\text{acc}_{\text{before\_attack}} - \text{acc}_{\text{after\_attack}})}{\text{acc}_{\text{before\_attack}}}$.
    In our work, ASR is calculated after applying adversarial detectors, which aim to purify the data by detecting and filtering out adversarial examples from the traffic.
    So, we replace \(\text{acc}_{\text{after\_attack}}\) with \(\text{acc}_{\text{purified}}\) to reflect the improved accuracy after purification. Thus, ASR in our experiments is calculated as: 
    $\text{ASR} = \frac{(\text{acc}_{\text{before\_attack}} - \text{acc}_{\text{purified}})}{\text{acc}_{\text{before\_attack}}}$.

     \textbf{Area Under the Receiver Operating Characteristic Curve (AUC-ROC):} 
     The AUC-ROC measures the adversarial detector's ability to distinguish between adversarial and clean examples~\cite{Manda9709532}. 
     It is computed by plotting the True Positive Rate (TPR) against the False Positive Rate (FPR) at various thresholds, where $\text{TPR} = \frac{\text{TP}}{\text{TP} + \text{FN}}$ 
     is defined by the proportion of actual adversarial examples that are correctly identified by the detector
     and $\text{FPR} = \frac{\text{FP}}{\text{FP} + \text{TN}}$ 
     is defined by the proportion of actual clean examples that are incorrectly flagged as adversarial.
     A higher AUC-ROC indicates better separation between adversarial and clean examples. 
     


\subsubsection{Adversarial Data Evaluation Metrics} \label{subsubsection:eval_adv_data}
As discussed in Section~\ref{lbl:quality_evaluation_of_generated_data}, a key method for evaluating the validity of generated adversarial data is to measure distributional differences from real data using statistical metrics. This section describes the distribution and feature-level metrics commonly used in the literature~\cite{schoen2022towards} and adopted in this study to assess whether \ND-generated data mimic legitimate network traffic. 

To satisfy \textit{Realism (Fidelity)}, we employed two statistical metrics: (i) \textit{First-order Wasserstein Distance}~\cite{pmlr-v97-wong19a}, which captures the minimal effort required to transform one distribution into another in a geometric sense. A low value indicates that adversarial flows lie close to the true data manifold. (ii) \textit{Maximum Mean Discrepancy (MMD)}~\cite{DBLP:journals/jmlr/GrettonBRSS12} measures the difference between distributions by comparing the means of their representations. We reported the biased estimate of squared MMD in a Reproducing Kernel Hilbert Space (RKHS). A low MMD value indicates that the generated and real distributions are similar.
Demonstrating these two scores close to the baseline can confirm that NetDiffuser's adversarial samples are effectively drawn from the same statistical distribution as real traffic. 

To satisfy \textit{Diversity (Fairness)}, we employed: (iii) \textit{Density vs. PCA Projection}, 
which visualizes kernel density estimates of real and adversarial samples along their first principal component to assess distributional alignment and provides a visual insight into how closely the distributions are aligned. 
(iv) \textit{Absolute difference between Pearson Correlation Heatmap}~\cite{TabDDPM10.5555/3618408.3619133} computes the absolute difference between real and adversarial feature‐correlation matrices. A low difference score indicates that the structural dependencies among features are preserved, supporting the claim that the generated data maintains diversity comparable to the original dataset. Together, these metrics will confirm that \ND~captures distribution-preserving inter-feature relationships.
%
To satisfy \textit{Originality (Authenticity)}, it is important to note that adversarial examples in \ND~are inherently constructed as minimal perturbations of real network flows. As shown in  Algorithm~\ref{alg:net_diffuser}, each reverse‐diffusion step begins from a noisy version of a real feature vector (Step~\ref{algo2:sampling} and  Step~\ref{algo2:fwd-loop-st}), ensuring that no \(x_{\mathrm{adv}}\) can exactly duplicate any training point. 

Finally, we performed a \textit{manual sanity check} by inspecting a subset of generated adversarial samples to verify that they exhibit plausible feature patterns and do not contain any abnormal or unrealistic values. Additional details and results are available on our GitHub repository.

\subsection{Adversarial Detectors}
\label{sec:defenses}
To assess the effectiveness of \ND, we compared its attack performance against two state-of-the-art defense mechanisms, specifically adversarial detectors: (i) MANDA~\cite{Manda9709532}, and (ii)   
Artifact~\cite{feinman2017detecting}.
\subsubsection{MANDA Detector}
MANDA (MANifold and Decision boundary-based Adversarial detection)~\cite{Manda9709532} was developed to detect adversarial examples in DL-based NIDS by exploiting inconsistencies between Manifold and Decision Boundary.
It has two detection modules: (i) the \textit{Manifold} detector, which flags inputs where the manifold evaluation is inconsistent with the NIDS model prediction, and (ii) the \textit{Decision Boundary}, which evaluates whether an input is a near-boundary example in high-dimensional space.
\subsubsection{Artifact Detector}
The Artifact~\cite{feinman2017detecting} detector identifies adversarial examples by analyzing statistical anomalies in the deep feature space of a neural network. It operates under the assumption that adversarial samples lie off the true data manifold and leverages two complementary signals for detection: (i) \textit{Bayesian uncertainty estimates} derived from Monte Carlo dropout, and (ii) \textit{Kernel density estimates} computed in the feature space of the last hidden layer. These metrics are then used to train a simple logistic regression classifier to distinguish adversarial examples from benign/noisy inputs.  Artifact was originally created for image datasets (e.g., MNIST and CIFAR-10). We adapted it to the NIDS domain.

\subsection{Results}
\label{sec:results}
This section evaluated the robustness of \ND~against baseline adversarial attacks and defenses. We conducted experiments using multiple anomaly detector architectures.
We reported performance using five key metrics: (1) Before-Attack accuracy, which reflects the baseline performance of the anomaly detector on clean, unperturbed inputs, (2) After-Attack accuracy, which represents the accuracy of anomaly detector after an adversarial attack,  (3) After-Purification accuracy, which denotes the accuracy after applying MANDA; (4) Attack Success Rate (ASR); and (5) AUC-ROC. 

\begin{table}[ht]
\caption{Comparison of AUC-ROC scores (lower is better) for Artifact and MANDA detectors under different adversarial attacks. 
Lower AUC-ROC values indicate reduced detector performance.}
\label{tab:auc_roc_artifact_manda}
\centering
\resizebox{\columnwidth}{!}{%
\begin{tabular}{l|cc|cc|cc}
\hline
\textbf{\makecell{Attack}} & \multicolumn{2}{c|}{\textbf{UNSW-NB15}} & \multicolumn{2}{c|}{\textbf{CICIDS2017}} & \multicolumn{2}{c}{\textbf{CICDDoS2019}} \\
\cline{2-7}
 & Artifact & MANDA & Artifact & MANDA & Artifact & MANDA \\
\hline
\multicolumn{7}{c}{Target model: \textbf{MLP-1L}} \\
\hline
FGSM           & 0.9985 &   0.9181     &    0.9902     &    0.8898    &   0.9410     &   0.9928     \\
\textit{ND-FGSM} & \textbf{0.5021} &    0.7579    &  \textbf{0.5254}     &   0.8579     &   \textbf{0.6324}    &  0.9221     \\
PGD             &   0.9996     &   0.9165     &   0.9945     &   0.8991     &   0.9155     &    0.9913    \\
\textit{ND-PGD}  &    \textbf{0.4770}    &   0.7578     &   \textbf{0.5373}     &   0.8556     &   \textbf{0.6662}     &   0.9347     \\
ACG            &   0.9553     &   0.9040     &   0.9980     &    0.9393    &   0.9666    &   0.9912     \\
\textit{ND-ACG}  &   \textbf{0.4656}     &   0.7203     &    \textbf{0.5259}    &    0.8866    &   \textbf{0.6520}    &   0.9717     \\

\hline
\end{tabular}
}
\end{table}

\subsubsection{\ND~(ND) Against Detectors}
\label{subsub:ASR_and_AUC}
To understand how \ND~affects adversarial detectors, we compare the performance of MANDA and Artifact under both baseline (FGSM, PGD, ACG) and \ND~attacks (ND-FGSM, ND-PGD, ND-ACG) using the MLP-1L model across all datasets. Table~\ref{tab:auc_roc_artifact_manda} shows the results. Artifact consistently outperforms MANDA on baseline attacks (rows for FGSM, PGD, and ACG) for UNSW-NB15 and CICIDS2017, while MANDA performs slightly better (higher AUC-ROC value) on CICDDoS2019.  
Interestingly, this contrasts with findings from the original MANDA paper, where MANDA achieved superior performance on CICIDS2017. Since our evaluation uses the same MLP-1L architecture, this discrepancy highlights how detector performance can be sensitive to differences in preprocessing, implementation details, or dataset handling.

However, Artifact’s performance against \ND~is significantly weaker, with AUC-ROC scores approaching the random guessing threshold ($\approx 0.50$) across all datasets. Moreover, Artifact does not provide mechanisms to mitigate the impact of adversarial examples on the NIDS once detected. Due to these limitations, we conduct the remainder of our evaluation using MANDA, which demonstrates more stable performance and also offers purification capability features that are essential for practical, real-world defense scenarios.

\subsubsection{\ND~Against MANDA}
This subsection presents the evaluation of \ND~against the MANDA detector. 

Tables~\ref{tab:unsw_evaluation},~\ref{tab:cicids_evaluation}, and~\ref{tab:cicddos_evaluation} summarize the anomaly detector’s performance. 
%
The anomaly detectors achieved strong \textit{Before-Attack} accuracy of approximately 89\% on the UNSW-NB15 dataset, around 95\% on the CICIDS2017 dataset, and up to 99\% on the CICDDoS2019 dataset across different model architectures.
After the attack is applied, detection performance drops significantly. Under FGSM, 
the accuracy of the MLP-1L model on the UNSW-NB15 dataset, as shown in Table~\ref{tab:unsw_evaluation}, drops from 89.12\% to 16.43\%, and the CNN-2L model from 89.14\% to 29.34\%. Meanwhile, the accuracy under \ND-FGSM remains substantially higher at 65.71\% (MLP-1L) and 61.62\% (CNN-2L), respectively. 
This difference arises because FGSM creates adversarial examples by applying perturbations that directly push the input toward a misclassification. These perturbations distort features in unnatural ways, leading to more severe misclassifications and a sharper drop in accuracy. In 
an NIDS environment, such abrupt and significant performance degradation draws attention to the system’s compromised state, potentially prompting further inspection by operators or automated monitoring tools. In contrast, the more moderate accuracy drop observed with \ND-FGSM allows adversarial traffic to blend in more seamlessly, making it less likely to raise suspicion while still compromising the system. We observed similar trends across the CICIDS2017 and CICDDoS2019 datasets as shown in Tables~\ref{tab:cicids_evaluation} and \ref{tab:cicddos_evaluation},
where \ND~consistently results in a more moderate accuracy drop compared to baseline attacks.

To mitigate such threats, adversarial detectors are often deployed as an added layer of defense. 
However, \ND-generated attacks pose a unique challenge. By causing only a moderate drop in accuracy, they avoid raising immediate suspicion. 
{\bf This subtle degradation in performance makes \ND-crafted adversarial traffic harder to flag as adversarial.} 



\begin{table}[ht]
\caption{Performance evaluation of anomaly detector models using MANDA 
on \textbf{UNSW-NB15}. 
(Lower AUC-ROC is better; higher ASR is better.)
}
\label{tab:unsw_evaluation}
\scriptsize
\centering

\begin{tabular}{l|cc|cc}
\hline
\textbf{\makecell{Attack}} & \textbf{\makecell{After-Attack\\Acc (\%)}} & \textbf{\makecell{After-Purif.\\Acc (\%)}} & \textbf{\makecell{ASR\\(\%)} 
} & \textbf{\makecell{AUC\\ROC} 
} \\
\hline

\multicolumn{5}{c}{\textbf{MLP-1L} (Before-Attack Accuracy: 89.12\%)} \\
\hline
FGSM            & 16.43 & 62.23 & 30.18 & 0.9181\\
\textit{ND-FGSM} & 65.71 & 46.84 & \textbf{47.45} &\textbf{0.7579}\\
PGD             & 16.43 & 63.72 & 28.50 & 0.9165\\
\textit{ND-PGD}  & 67.48 & 48.32 & \textbf{45.78} & \textbf{0.7578} \\
ACG             & 31.35 & 59.04 & 33.75 & 0.9040\\
\textit{ND-ACG}  & 71.01 & 46.62 & \textbf{47.68} & \textbf{0.7203}\\
\hline
\multicolumn{5}{c}{\textbf{MLP-5L} (Before-Attack Accuracy: 88.29\%)} \\
\hline
FGSM            & 35.32 & 60.68 & 31.27 &0.9144\\
\textit{ND-FGSM} & 70.21 & 44.36 & \textbf{49.76} & \textbf{0.6868}\\
PGD             & 17.92 & 67.67 & 23.35 & 0.9228\\
\textit{ND-PGD}  & 73.78 & 45.55 & \textbf{48.40} & \textbf{0.6562}\\
ACG             & 37.48 & 55.88 & 36.71 & 0.8826\\
\textit{ND-ACG} & 75.65 & 45.80 & \textbf{48.12} & \textbf{0.6408}\\
\hline
\multicolumn{5}{c}{\textbf{CNN-2L} (Before-Attack Accuracy: 89.14\%)} \\
\hline
FGSM            & 29.34 & 73.67 & 17.36 & 0.9394 \\
\textit{ND-FGSM} & 61.62 & 55.69 & \textbf{37.53} & \textbf{0.7756}\\
PGD             & 11.82 & 63.72 & 28.51 & 0.9165\\
\textit{ND-PGD}  & 62.86 & 59.87 & \textbf{32.84} & \textbf{0.8309}\\
ACG             & 27.26 & 66.75 & 25.11 & 0.9344\\
\textit{ND-ACG}  & 70.10 & 59.40 & \textbf{33.37} & \textbf{0.7859}\\
\hline
\multicolumn{5}{c}{\textbf{CNN-3L} (Before-Attack Accuracy: 87.65\%)} \\
\hline
FGSM            & 42.38 & 61.44 & 29.90 & 0.8867\\
\textit{ND-FGSM} & 70.51 & 51.07 & \textbf{41.73} & \textbf{0.7910}\\
PGD             & 12.51 & 64.69 & 26.20 & 0.9256\\
\textit{ND-PGD}  & 71.12 & 48.67 & \textbf{44.47} & \textbf{0.7320}\\
ACG             & 23.09 & 66.33 & 24.33 & 0.9256\\
\textit{ND-ACG}  & 73.03 & 45.86 & \textbf{47.68} & \textbf{0.6969}\\
\hline
\end{tabular}
\end{table}


\begin{table}[ht]
\caption{Performance evaluation of anomaly detector models using MANDA on \textbf{CICIDS2017} dataset. 
(Lower AUC-ROC is better; higher ASR is better.)
}
\label{tab:cicids_evaluation}
\scriptsize
\centering

\begin{tabular}{l|cc|cc}
\hline
\textbf{\makecell{Attack}} & \textbf{\makecell{After-Attack\\Acc (\%)}} & \textbf{\makecell{After-Purif.\\Acc (\%)}} & \textbf{\makecell{ASR\\(\%)} 
} & \textbf{\makecell{AUC\\ROC} 
} \\
\hline
\multicolumn{5}{c}{\textbf{MLP-1L} (Before-Attack Accuracy: 95.30\%)} \\
\hline
FGSM            & 5.59 & 70.96 & 25.54 & 0.8898 \\
\textit{ND-FGSM} & 70.10 & 70.29 & \textbf{26.24} & \textbf{0.8579}\\
PGD             & 5.59 & 70.14 & 26.40 & 0.8991\\
\textit{ND-PGD}  & 71.98 & 69.85 & \textbf{26.71} & \textbf{0.8556}\\
ACG             & 49.58 & 78.93 & 17.18 & 0.9393\\
\textit{ND-ACG}  & 68.76 & 72.72 & \textbf{23.70} & \textbf{0.8866} \\
\hline
\multicolumn{5}{c}{\textbf{MLP-5L} (Before-Attack Accuracy: 96.51\%)} \\
\hline
FGSM            & 13.35 & 75.68 & 21.59 & 0.9132 \\
\textit{ND-FGSM} & 63.19 & 71.66 & \textbf{25.74} & \textbf{0.8616}\\
PGD             & 5.39 & 76.51 & 20.72 &0.9105 \\
\textit{ND-PGD}  & 59.38  & 74.81 & \textbf{22.48} & \textbf{0.8898}\\
ACG             & 49.07 & 78.77 & 18.38 & 0.9291\\
\textit{ND-ACG}  & 70.59 & 78.65 & \textbf{18.51} & \textbf{0.8669} \\
\hline
\multicolumn{5}{c}{\textbf{CNN-2L} (Before-Attack Accuracy: 95.53\%)} \\
\hline
FGSM            & 5.32 & 73.98 & 22.56 &0.9127\\
\textit{ND-FGSM} & 73.84 & 63.50 & \textbf{33.53} & \textbf{0.7648}\\
PGD             & 4.47 & 75.55 & 20.91 & 0.9048\\
\textit{ND-PGD}  & 78.19 & 60.43 & \textbf{36.74} & \textbf{0.7303}\\
ACG             & 5.57 & 75.90 & 20.55 & 0.9232\\
\textit{ND-ACG}  & 85.99 & 67.24 & \textbf{29.61} & \textbf{0.7082}\\
\hline
\multicolumn{5}{c}{\textbf{CNN-3L} (Before-Attack Accuracy: 96.52\%)} \\
\hline
FGSM            & 28.09 & 72.52 & 24.87 & 0.8433\\
\textit{ND-FGSM} & 79.30 & 71.69 & \textbf{25.72} & \textbf{0.7809} \\
PGD             & 3.49 & 79.55 & 17.59 & 0.9236 \\
\textit{ND-PGD}  & 84.10 & 70.91 & \textbf{25.42} & \textbf{0.7516} \\
ACG             & 5.13 & 75.39 & 21.90 &0.9124 \\
\textit{ND-ACG}  & 85.35 & 72.54 & \textbf{24.84} & \textbf{0.7553}\\
\hline
\end{tabular}
\end{table}


\begin{table}[ht]
\caption{Performance evaluation of anomaly detector models using MANDA 
on \textbf{CICDDoS2019}. 
(Lower AUC-ROC is better; higher ASR is better.)
}
\label{tab:cicddos_evaluation}
\scriptsize
\centering

\begin{tabular}{l|cc|cc}
\hline
\textbf{\makecell{Attack}} & \textbf{\makecell{After-Attack\\Acc (\%)}} & \textbf{\makecell{After-Purif.\\Acc (\%)}} & \textbf{\makecell{ASR\\(\%)}} & \textbf{\makecell{AUC\\ROC}  } \\
\hline
\multicolumn{5}{c}{\textbf{MLP-1L} (Before-Attack Accuracy: 99.77\%)} \\
\hline
FGSM            & 1.27 & 84.19 & 15.61 & 0.9928\\
\textit{ND-FGSM} & 94.91 & 82.80 & \textbf{17.01} & \textbf{0.9221} \\
PGD             & 1.27 & 84.15 & 15.65 & 0.9913 \\
\textit{ND-PGD}  & 70.51 & 79.78 & \textbf{20.04} & \textbf{0.9347}\\
ACG             & 45.04 & 92.47 & 7.32 & 0.9912 \\
\textit{ND-ACG}  & 76.61  & 91.98 & \textbf{7.81} & \textbf{0.9717} \\
\hline
\multicolumn{5}{c}{\textbf{MLP-5L} (Before-Attack Accuracy: 99.82\%)} \\
\hline
FGSM            & 66.04 & 94.25 & 5.58 & 0.9908\\
\textit{ND-FGSM} & 97.39 & 64.37 & \textbf{35.51} & \textbf{0.7851} \\
PGD             & 44.47 & 97.38 & 2.44 & 0.9941\\
\textit{ND-PGD}  & 96.44 & 72.37 & \textbf{27.50} & \textbf{0.8458}\\
ACG             & 50.35 & 97.88 & 1.94 & 0.9952 \\
\textit{ND-ACG}  & 96.23 & 72.42 & \textbf{27.45} & \textbf{0.8600}\\
\hline
\multicolumn{5}{c}{\textbf{CNN-2L} (Before-Attack Accuracy: 99.75\%)} \\
\hline
FGSM            & 43.59 & 98.33 & 1.43 & 0.9976\\
\textit{ND-FGSM} & 92.58 & 88.20 & \textbf{11.58} & \textbf{0.9988}\\
PGD             & 5.41 & 98.54 & 1.22 & 0.9969\\
\textit{ND-PGD}  & 89.24 & 88.96 & \textbf{10.82} & \textbf{0.9510}\\
ACG             & 28.68 & 98.54 & 1.22 & 0.9988\\
\textit{ND-ACG}  & 95.74 & 68.53 & \textbf{31.30} & \textbf{0.8277}\\
\hline
\multicolumn{5}{c}{\textbf{CNN-3L} (Before-Attack Accuracy: 99.87\%)} \\
\hline
FGSM            & 51.26 & 97.26 & 2.62 & 0.9911\\
\textit{ND-FGSM} & 91.05 & 71.78 & \textbf{28.12} & \textbf{0.8566}\\
PGD             & 1.38 & 97.23 & 2.64 & 0.9951\\
\textit{ND-PGD}  & 85.70 & 81.03 & \textbf{18.87} & \textbf{0.9272} \\
ACG             & 9.32 & 98.03 & 1.85 & 0.9966\\
\textit{ND-ACG}  & 95.80 & 68.28 & \textbf{31.63} & \textbf{0.8119}\\
\hline
\end{tabular}
\end{table}

To evaluate whether such stealthy examples can evade detection, we assessed MANDA’s ability to identify and filter \ND-generated inputs.
We measured how effectively each detector performs when deployed in front of the anomaly detection model, as illustrated in Figure~\ref{fig:DL-NIDS}.
The defense module filters inputs by identifying and removing adversarial examples, and only the remaining traffic (referred to as purified traffic) is passed to the anomaly detector model. 
The higher \textit{After-Purification} accuracy value indicates that the defense successfully filters out the poisoned instances while retaining the un-poisoned ones.
For example, in the UNSW-NB15 dataset, as shown in Table~\ref{tab:unsw_evaluation}, applying the MANDA detector to FGSM-perturbed inputs for the MLP-1L model improves the accuracy from 16.43\% to 62.23\%, and for the CNN-2L model from 29.34\% to 73.67\%. However, when evaluating the same defense against \ND-FGSM inputs, the recovery is significantly less effective. For the MLP-1L, \textit{After-Purification} accuracy drops from 65.71\% to 46.84\%, and for CNN-2L, from 61.62\% to 55.69\%. This suggests that the detector fails to identify and filter out the perturbations introduced by \ND. A similar pattern is observed across other datasets. 
For instance, on the CICIDS2017 dataset, as shown in Table~\ref{tab:cicids_evaluation}, the purified accuracy of the MLP-1L model improves significantly from 5.59\% to 70.14\% after applying MANDA to PGD-perturbed inputs. In contrast, for the \ND-PGD-perturbed inputs, the accuracy slightly decreases from 71.98\% to 69.85\%, highlighting the subtlety of \ND's perturbations.
Similarly, on the CICDDoS2019 dataset (Table~\ref{tab:cicddos_evaluation}), the purified accuracy increases significantly from 1.27\% to 84.19\% for FGSM attack, but drops from 94.91\% to 82.80\% for \ND-FGSM attack.

While \ND-generated adversarial examples frequently bypass defenses, resulting in reduced \textit{After-Purification} accuracy, we also observe cases where purification leads to improved performance. 
For instance, on the CICIDS2017 dataset with MLP-1L, applying the defense/detector increases the accuracy for \ND-ACG attack from 69.76\% to 72.72\%.
However, these improvements are typically small and inconsistent and remain well below the clean baseline accuracy of 95.30\%. Although MANDA improves classification accuracy on both baseline and \ND~attacks, the gains for \ND~are often small and inconsistent.
This limited recovery highlights a key issue: 
\ND-generated adversarial examples evade detection by avoiding the sharp performance drops that typically signal suspicious activity.
They also remain resilient to purification efforts, making them difficult to eliminate once deployed in the DL-based NIDS environment with defense.
\textbf{Adversarial Attack Performance:} We evaluated the adversarial attack performance using two metrics (defined in Section~\ref{subsubsection:eval_model}): 
ASR, which measures the percentage of successfully misclassified samples, and \textit{AUC-ROC}, which captures the anomaly detector's ability to distinguish between clean and adversarial examples. 
The baseline attacks \textit{i.e.,} FGSM, PGD, and ACG degrade anomaly detector performance by significantly lowering the \textit{After-Attack} accuracy, primarily due to noticeable shifts in the feature distribution. In contrast, \ND~introduces natural perturbations that remain better aligned with the original distribution, causing subtle changes in feature space (as discussed in Section~\ref{subsubsection:stat_distance}).

Across all datasets and models, \ND~consistently achieved a higher ASR on MANDA than the baseline attacks.
For example, on the UNSW-NB15 dataset with the MLP-1L model, as shown in Table~\ref{tab:unsw_evaluation}, \ND-FGSM achieves a higher ASR of 47.45\%, compared to 30.18\% for FGSM. 
Similarly, with the CNN-2L model, \ND-FGSM reaches 37.53\%, outperforming FGSM’s 17.36\%.
Similar trends are observed across all attack types--\ND-crafted adversarial examples consistently result in more successful evasion.

Concurrently, \ND-generated adversarial examples yielded lower AUC-ROC scores, as computed by the MANDA detector, indicating its reduced effectiveness in differentiating between adversarial and clean examples.
For instance, while FGSM on the MLP-1L achieves an AUC-ROC of 0.9181, \ND-FGSM reduces it significantly to 0.7579 (Table~\ref{tab:unsw_evaluation}). This trend was consistent across all datasets, with \ND~generated examples consistently yielding lower AUC-ROC values than the baseline attacks. 
However, in some cases, such as MLP-5L with ACG on CICIDS2017 and MLP-1L with ACG on CICDDoS2019, the ASR of \ND-ACG is comparable to that of the baseline ACG.
Despite the similarity in ASR, the AUC-ROC scores for \ND-ACG remain lower, indicating that \ND~continues to effectively induce misclassifications while the detector still struggles to distinguish these examples from clean inputs. However, \ND~adds a significant runtime overhead compared to baseline attacks, which is due to the expected cost of the multi-step diffusion process.
\subsubsection{Assessment of Adversarial Data using Statistical Metric} \label{subsubsection:stat_distance}
To evaluate how closely adversarial examples resemble the original data distribution, we used two statistical measures: \textit{Maximum Mean Discrepancy (MMD)}~\cite{DBLP:journals/jmlr/GrettonBRSS12} and \textit{Wasserstein Distance}~\cite{pmlr-v97-wong19a}. Unlike norm-constrained attacks, which are typically evaluated using perturbation-based metrics such as $L_0$, $L_2$, or $L_\infty$ norms, our approach generated Natural Adversarial Examples (NAEs) that are unconstrained in form but statistically aligned with the original data. 
Consequently, we adopted distribution-level metrics to assess how closely the generated examples preserve the global structure of the original data distribution.

\begin{table}[ht]
\caption{Comparison of Wasserstein and Maximum Mean Discrepancy (MMD) distances (closer to 0 is better) between original and adversarial samples. 
}
\label{tab:stat_dist_main}
\scriptsize
\centering
\begin{tabular}{l|cc|cc|cc}
\hline
\textbf{\makecell{Attack}} & \multicolumn{2}{c|}{\textbf{UNSW-NB15}} & \multicolumn{2}{c|}{\textbf{CICIDS2017}} & \multicolumn{2}{c}{\textbf{CICDDoS2019}} \\
\cline{2-7}
& Wass. & MMD 
& Wass. & MMD 
& Wass. & MMD 
\\
\hline
\multicolumn{7}{c}{Target model: \textbf{MLP-1L}} \\
\hline
FGSM            & 0.1519 & 0.4904 & 0.1466 & 0.5931 & 0.1464 & 0.6319 \\
\textit{ND-FGSM} & 0.0187 & \textbf{0.0043} & 0.0182 & \textbf{0.0363} & 0.0291 & \textbf{0.1726} \\
PGD             & 0.1106 & 0.4055 & 0.1357 & 0.6528 & 0.1300 & 0.7588 \\
\textit{ND-PGD}  & 0.0170 & \textbf{0.0041} & 0.0182 & \textbf{0.0330} & 0.0319 & \textbf{0.1231} \\
ACG             & 0.1179 & 0.2852 & 0.1456 & 0.4884 & 0.1328 & 0.3874 \\
\textit{ND-ACG}  & 0.0141 & \textbf{0.0075} & 0.0142 & \textbf{0.0083} & 0.0309 & \textbf{0.0607} \\
\hline
\multicolumn{7}{c}{Target model: \textbf{CNN-2L}} \\
\hline
FGSM            & 0.1379 & 0.3599 & 0.1366 & 0.5185 & 0.1564 & 0.4844 \\
\textit{ND-FGSM} & 0.0436 & \textbf{0.0817} & 0.0138 & \textbf{0.0038} & 0.1050 & \textbf{0.1266} \\
PGD             & 0.0653 & 0.1603 & 0.0617 & 0.2554 & 0.0366 & 0.1335 \\
\textit{ND-PGD}  & 0.0297 & \textbf{0.0585} & 0.0125 & \textbf{0.0058} & 0.1025 & \textbf{0.0964} \\
ACG             & 0.1087 & 0.2201 & 0.0927 & 0.2724 & 0.0855 & 0.2011 \\
\textit{ND-ACG}  & 0.0190 & \textbf{0.0117} & 0.0103 & \textbf{0.0011} & 0.0238 & \textbf{0.0725} \\
\hline
\end{tabular}
\end{table}

Table~\ref{tab:stat_dist_main} presents the MMD and Wasserstein values between clean and adversarial examples across all the attack types, models, and datasets. In all the cases, \ND~crafted adversarial examples exhibited lower MMD compared to the baseline attacks, which indicates higher similarity between adversarial and original distribution. We observed a similar trend for Wasserstein distance in most settings. 
However, in some cases, such as with CNN-2L and CNN-3L on the CICDDoS2019 dataset, the Wasserstein distance for PGD is slightly lower than that for \ND-PGD-generated adversarial examples, despite the MMD being lower for \ND-PGD.


\begin{figure}[htbp]
    \centering
    \begin{subfigure}[t]{1.0\linewidth}
        \centering
        \includegraphics[width=\linewidth]{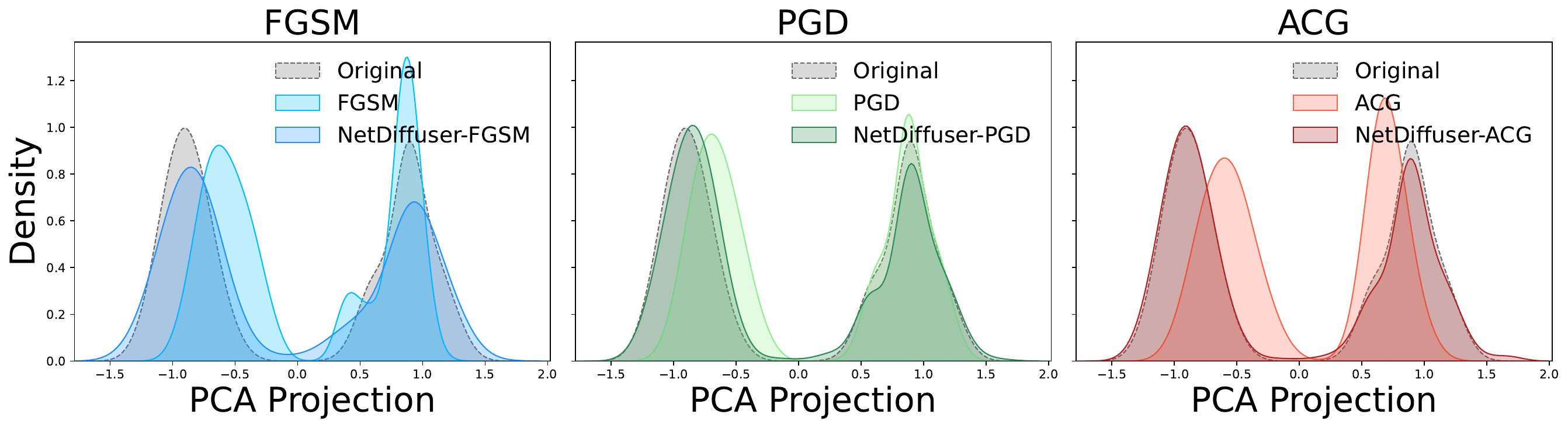}
        \caption{UNSW-NB15}
        \label{fig:unsw-pca-mlp1}
    \end{subfigure}
    
    \vspace{1em}

    \begin{subfigure}[t]{1.0\linewidth}
        \centering
        \includegraphics[width=\linewidth]{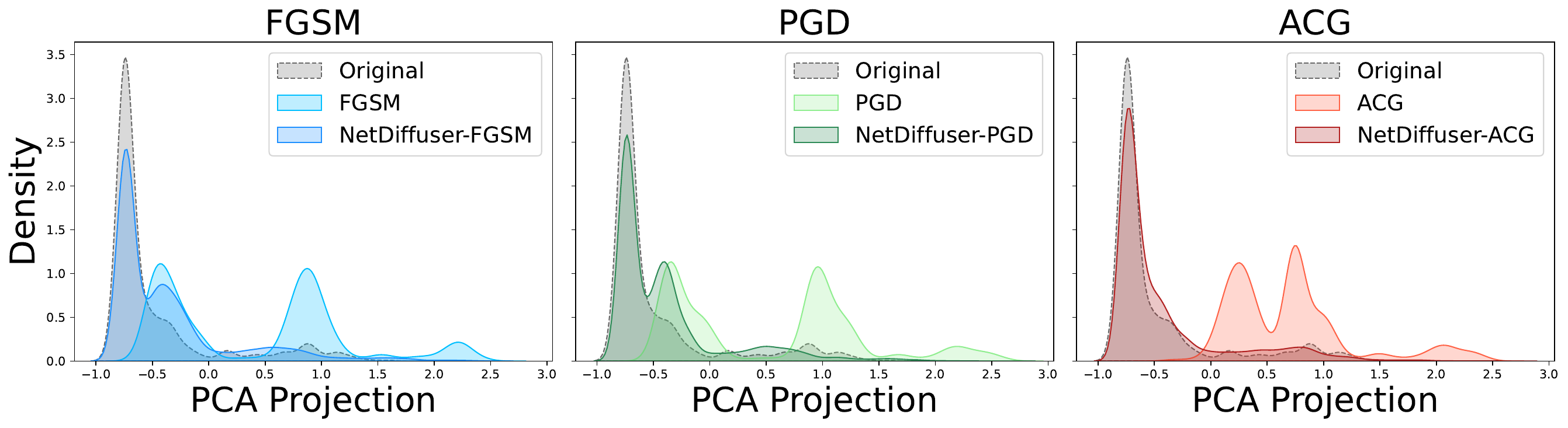}
        \caption{CICIDS2017}
        \label{fig:cicids-pca-mlp1}
    \end{subfigure}

    \begin{subfigure}[t]{1.0\linewidth}
        \centering
        \includegraphics[width=\linewidth]{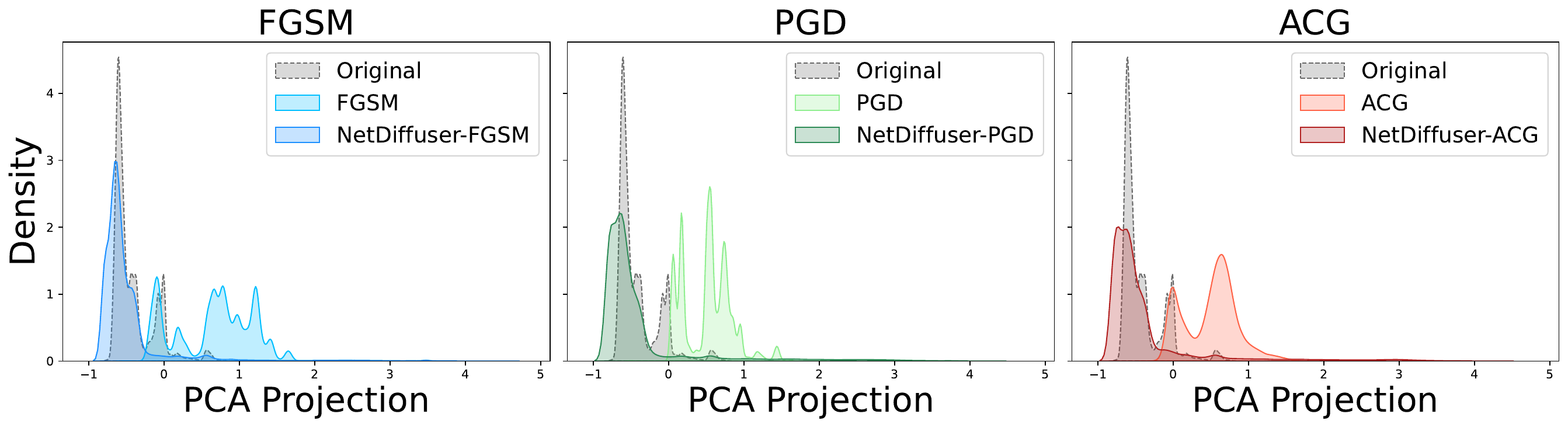}
        \caption{CICDDoS2019}
        \label{fig:cicddos-pca-mlp1}
    \end{subfigure}

    \caption{PCA density plots comparing clean and adversarial samples targeting the MLP-1L model across UNSW-NB15, CICIDS2017, and CICDDoS2019 datasets.}
    \label{fig:pca-combined-mlp}
    \vspace{-0.1in}
\end{figure}

To illustrate the closeness, we plotted the PCA-based density distribution of adversarial and clean examples in Figure~\ref{fig:pca-combined-mlp}, which shows the distribution for adversarial examples targeting the MLP-1L model. 
Adversarial examples generated by \ND~tend to cluster more closely with clean data in the projected PCA space. This visual overlap was consistent with the lower MMD and Wasserstein scores, suggesting that \ND-crafted adversarial examples remain statistically closer to the original data distribution.
In contrast, baseline attacks exhibit a noticeable divergence from clean data, explaining their higher statistical distances and more abrupt perturbations. 

\begin{figure}[htbp]
    \centerline{\includegraphics[width=0.5\textwidth]{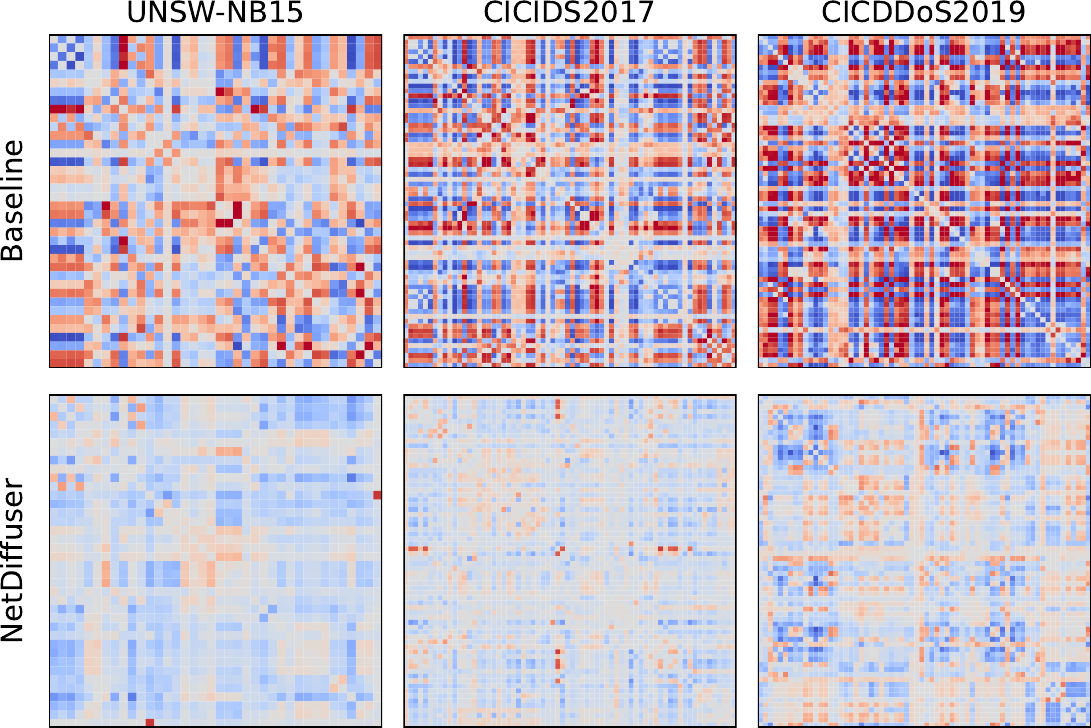}}
    \caption{The comparison of absolute difference between Pearson's correlation coefficient computed on original data and adversarial data generated by PGD and \ND-PGD on MLP-1L. Intense red indicates stronger correlations in the original data; intense blue indicates stronger correlations in the adversarial data. \ND~better preserves the correlation structure, as shown by lighter colors. 
    }
    \label{fig:correlation}
\end{figure}

We visualized the feature-wise correlation in Figure~\ref{fig:correlation}.
\ND-generated examples introduce significantly smaller deviations in the correlation structure compared to the baseline PGD attack. This showed that \ND~closely mimicked the structure of network traffic, which is crucial for maintaining the semantic consistency of the flows and avoiding detection. Additional plots and results for MLP-5L and CNN-3L models are available in our GitHub repository.

%% file: conclusion.tex
\section{Conclusion} \label{sec:conclusion}
We introduced \ND, a novel method for crafting adversarial examples that adhere to domain-specific constraints, ensuring realistic and valid network traffic. 
\ND~pioneers the use of NAEs
in the NIDS domain, leveraging diffusion models to create imperceptible perturbations. Our experimental results demonstrate the effectiveness of \ND, achieving up to a 29.93\% higher attack success rate and reduced AE detection performance by up to 0.267 and 0.534 in AUC-ROC score for the MANDA and Artifact detectors, respectively. These findings underscore the need for robust defenses that can effectively address both traditional and naturally occurring adversarial behaviors in NIDS. 
\ND~can improve attacks' stealth and overall effectiveness, but at the expense of extra runtime overhead.
Future work may focus on streamlining the sampler or hybridizing diffusion with faster methods to alleviate computational overhead.  
Future research will extend the proposed \ND~to gray-box and black-box settings and devise robust defense mechanisms capable of mitigating both traditional adversarial examples and NAEs. 

\section*{Acknowledgment}
This work was sponsored by the DEVCOM Analysis Center and was accomplished under Cooperative Agreement Numbers W911QX23D0009 and W911NF2220001. This work was also partially funded by the National Science Foundation (NSF) awards CNS-2148358 and OIA-2417062.